\documentstyle[12pt]{article}

\def\eop{\vspace*{\fill}\pagebreak}

\begin{document}

\begin{titlepage}
{\bf APRIL, 1993}\hfill       {\bf PUPT-1394}\\
\begin{center}

{\bf A FEW PROJECTS IN STRING THEORY }

\vspace{1.5cm}

{\bf  A.M.~Polyakov}

\vspace{1.0cm}

{\it  Physics Department, Princeton University,\\
Jadwin Hall, Princeton, NJ 08544-1000.\\
E-mail: polyakov@puhep1.princeton.edu}

\vspace{1.9cm}
\end{center}

\abstract{
In these lectures I discuss various unsolved problems of string
theory and
their relations to quantum gravity, 3d Ising model, large N QCD, and
quantum
cosmology. No solutions are presented but some new and perhaps useful
approaches are suggested.
}
\vfill
\end{titlepage}


  \section{Introduction}
I consider string theory as a set of unifying concepts and methods
in physics.
Technically speaking it consists of conformal field theories coupled
to the
self induced two dimensional gravitational field. Physically, it
describes
random surfaces with  different extra degrees of freedom, immersed
into
different ambient spaces. \par Such objects are frequent in nature -
phase
boundaries in critical phenomena or the surfaces of biological
membranes are
 notable examples. \par In a somewhat more subtle way  random
surfaces
appear in confining gauge theories. Here the lines of the color -
electric flux
form a closed string. The transition amplitudes, which are given by
 sums
over histories, are expressed now in terms of  random surfaces,
formed by
 propagating strings. \par Probably the most fascinating
application of the
string ideas is the theory of many dimensional gravity. Here one
views a
graviton as a tiny closed string. Under certain conditions this
approach gives
the only consistent theory of quantum gravity known today. \par There
are
several other applications of the techniques of string theory. For
example,
conformal field theory describes most of  two- dimensional
statistical
physics, the Kondo problem, polymers branching and perhaps  two
dimensional
turbulence. \par  It is important, when working with strings, to keep
in mind
all these complementary aspects of the subject.\par In these lectures
I
describe several projects concerning different parts of physics but
united by
their relations to strings.  My choices are determined  by personal
tastes and
there is no pretence for ``objectivity" whatever it means. Even within
this
limited perspective I omited the fascinating subject of conformal
turbulence,
since  I have discussed it recently elsewhere.  I  concentrate on the
open
problems which are rarely considered in the literature.

\section{ Quantum Gravity }
 In the conventional theory of quantum gravity one begins with the
notion of
continuous and curved space-time. One believes, that there exists
a
fluctuating metric tensor $ G_{\mu \nu}\left( x \right)$  which in
the vacuum
has non-zero Minkowskian expectation value. \par In the classical
limit this
value satisfies the Einstein equations. The size of the quantum
corrections at
the energy $ E$ is of the order of $ {E^{2} \over M^{2}}$  where $ M$
is the
Planck mass. Let us critically discuss this situation. \par First of
all, should
 space- time be continuous? It seems to me that the answer is no.
Imagine
the world which consists of the tangled network of world-lines.
Events in this
world consist of the crossing points of these lines. Our measuring of
time is
based on counting the number of crossings, separating two given
events. If so,
the lapse of the time, when ``nothing happens" is a meaningless
notion.\par So,
most probably, the continuity  of space-time is just an approximation
similar
to the one we use in condensed matter or hydrodynamics.\par Of
course, just the
idea of discreteness, even together with the correspondence principle
is not
enough to build the theory. It is somewhat ironic that  the
discreteness, while
being the most elementary of all possible notions, makes the theory
apparently
intractable. That is why we will concentrate on the continuous
theories. Let us
notice, however, that in two dimensions there exists an attractive
discrete
theory of gravity based on the matrix models, with the correct
continuous
limit.\par Passing to the continuous theory let us discuss if we can
use the
Einstein action as a starting point for the quantum theory. The
problem here is
non-renormalizability: the quantum corrections, diverge at  very high
energies
(and the energies of the virtual particles can be arbitrary
high).\par A
possible way out is to conjecture that the coupling constant in
gravity is
scale dependent and tends to zero at small distances. This
``antiscreening" of
the gravitational interaction is quite natural since the larger is
the cloud of
the virtual particles, the stronger the gravitational force is.\par
It is
instructive to compare the situation with the one in other
non-renormalizible theories. Let us consider a non-linear $ \sigma$ -
model

\begin{equation}
L\,=\, {1 \over 2\kappa_{0}}\,\left( \partial_{\mu}\vec{n}
\right)^{2}\,,\,\vec{n}^{2}=1\,,\,\vec{n}=\left(
\sqrt{1-\vec{\pi}^{2}}\,
,\vec{\pi}
 \right)
 \end{equation}
The $ \pi$ - meson field is the analogue of the  gravitation. In four
dimensions, perturbative interaction of pions as well as of the
gravitons leads
to the scale-dependence of the effective coupling  $ \kappa\left(
p\right)$ :

\begin{equation}
{\kappa\left( p \right) \over \kappa _{0}}\,=\,1+const\,\left(
\kappa_{0}
p^{2}\right)\,+\,\cdots
\end{equation}
 (where $ p$ is the momentum) formally similar to gravity.\par The
theory has a
phase transition at some coupling. In order to reach the continuum
limit, one
has to force the coupling $ \kappa_{0}$ to be close to the critical
value. In
this case the theory depends on one mass scale       $ M$ and the
effective
interaction behaves as:

\begin{equation}
\kappa\left( p^{2} \right)\,\infty\,p^{-2}
 \end{equation}
at ultra high energy $ \left( p^{2} >>M^{2}\right)$ while in the low
energy
limit we simply have:
\begin{equation}
\kappa\,\sim\,M^{-2}
\end{equation}
To be precise, there are also logarithmic corrections to these
formulas which
we disregarded. \par The physics in the above two regions is very
different.
Consider an expectation value of the field $ \vec{n}$. If we examine
a domain
of size smaller than the correlation length we find that the
broken and
unbroken phases are indistinguishable, while at low energies we
have pions
due to the symmetry breaking.

 \section{Phases in gravity}
Let us return to gravity. Can we have a similar picture? Is it
possible, that
the effective Newton constant behaves as (3)? What is the meaning of
the
unbroken phase in gravity? There are no ready answers to these
questions. I
shall only make a few comments on them.\par The problem with the
running Newton
constant is essentially that we have no calculational procedures to
deal with
the Einstein action. Some hope here may lie in the fact, that the
structure of
the operator algebra should be simplified due to the general
covariance. It
might consist of the few primary operators, plus infinite sets of
their
deformations caused by the diffeomorphisms. This is the picture we
have in
conformal field theories in which such a structure allows to obtain
exact
solutions. In the Einstein theory we are quite far from that, but it
is
worthwhile to keep in mind this possibility. It seems plausible that
general
covariance drastically simplifies any field theory.\par The question of
the
unbroken phase in gravity presents not only computational, but also
some
conceptual difficulties. \par In the most naive form, we would say
that in the
broken phase:
\begin{equation}
<G_{\mu\nu}> \, =\, \delta_{\mu\nu}
 \end{equation}
 while in the unbroken phase:
\begin{equation}
<G_{\mu\nu}>\,=\,0
 \end{equation}
In some sense (5) means that we have non-zero density of the
``space-time
substance".\par The problem with this definition is that (5) is a
gauge-dependent statement. It would be nice to have some gauge
invariant
criterion. One possibility is to consider a quantity:

\begin{equation}
A\left( \mu \right)=<exp-\mu \oint _{C}\sqrt{G_{\mu\nu} \left( x
\right)dx^{\mu}dx^{\nu}}>
\end{equation}
This represents a particle forming a closed loop in our manifold.
This
amplitude is independent of the shape of the loop $ C$ since we
average with
respect to all possible geometries. \par The test of the geometry can
be
performed by sending $ \mu\rightarrow\infty$.  In order to avoid some
trivial
renormalization problems it is more convenient to look at the
equivalent
object, which is a Laplace transform of (7) :
\begin{equation}
B\left( L \right)=<exp-{{\int_{0}}^{L}}\,G_{\mu\nu}\left( x\left( s
\right)
\right) \dot{x}^{\mu}\left( s \right)\dot{x}^{\nu}\left( s \right)ds>
 \end{equation}
This quantity is some analogue of the Wilson loops in gauge theory.
Its $ L$ -
independence at small $ L$ measures  effective dimensionality of the
space-
time, which is of course a scale dependent notion. \par If the
metric is flat,
$ B\left( L \right)$ has the standard, ``heat kernel" behavior:
\begin{equation}
B\left( L \right) \,\sim\,L^
{-D \over 2}
\end{equation}

On the other hand in the regime (6) metric $ G$ and $ \lambda G$ may
enter with
the same weight (the Weyl symmetry being unbroken) in which case we
expect:
\begin{equation}
B\left( L \right)\,\sim\,constant
\end{equation}
This means that the effective dimensionality is zero. Of course,
there could
exist phases with  intermediate dimensionality.\par
Unfortunately, in
general we don't have methods to study such questions. However this
can be done
in the case of 2d gravity.
\section {Self-tuning universe}
There is another  serious problem with quantum gravity. It seems that
we need a
hyperfine tuning for the cosmological constant. Actually it is even
worse than
that, we know that very many different constants in physics conspire
so as to
allow a non-trivial universe. It takes a tiny change in the fundamental
constants
to stop nucleosynthesis, extinguish the sun etc. See recent
discussion in
[1].\par All that may imply that the fundamental constants are in
fact some
self-tuning fields, and the really good universe is an attractor. At
the
present level of our knowledge, the best we can do is to play with
some toy
models which have self-tuning phenomenon.\par Today we know several
different
mechanisms for self-tuning. One of them is based on the idea of
uncontrollable
emissions of tiny ``baby-universes" [2]. This mechanism has been
extensively
discussed in the literature. The problem with it is the absence of
reasons for
the locality of coupling of the baby universes to the main one. Still
it is an
interesting possibility.\par Another idea, which may work for  the
cosmological
constant is the following [3]. The cosmological term in the Einstein
action
does not contain any derivatives of $ G_{\mu\nu}$ . In this case one
would
expect strong infrared interaction, at distances much larger than
the Planck
length.\par For example, if one considers conformal fluctuations of
the
metric,$G_{\mu\nu}\,=\,\varphi^{2}\delta_{\mu\nu}$, the cosmological
term is just
a $ \varphi^{4}$ - interaction.\par Such interactions tend to be
self-screened,
due to the ``zero charge" phenomenon. Thus starting from the ``bare"
cosmological
constant $ \Lambda_{0}$ one gets the physical $ \Lambda\rightarrow0$.
In the
``logarithmic" approximation, one gets:
\begin{equation}
\Lambda\sim{1 \over \log{MR}}\approx 0
 \end{equation}
 where $ M$ is the Planck mass and $ R$ is the inverse local
curvature.
Unfortunately this approximation is not adequate, but the ``infrared
screening"
idea may be valid anyway. As we will discuss later, one should add
the dilaton
field and consider infrared graviton - dilaton fluctuations together.
This has
not been done yet. \par A third self-tuning mechanism appears in the
$ c=1$
models for string theory. In this case one has an infinite tower of
tensor
fields $ B_{\mu 1} \ldots\mu_{N} \left( x \right) $ representing
higher string
excitations.\par However, string gauge invariance leads to the
following. The most
part of $ B\left( x \right)$ is just a gauge artefact. By a gauge
transformation one can replace the functions $ B_{\mu 1}\ldots
\mu_{N}\left( x
\right)$ by the set of constants $ b_{\mu 1}\ldots \mu_{N}$. Thus,
instead of
the  full-fledged fields, in this case we will be dealing with the
set of
varying coupling constants $ b_{\mu1}\ldots\mu_{N }$ which are just
the
remnants of $ B-s$ . A simple example of this phenomenon occurs with
the
two-dimensional vector quanta. They are described by the polarization
$
A_{\mu}\left(p \right)$, satisfying Lorentz condition: \par$
p_{\mu}A_{\mu}\left( p \right)\,=\,0$\par and defined modulo gauge
transformations:\par $ A_{\mu}\left( p\right)\rightarrow
A_{\mu}\left( p
\right)\,+\,p_{\mu}\phi\left( p \right)$ \par For $ p\neq0$ no states
remain, but
for $ p=0 $ the gauge transformation and the constraint disappear,
leaving us
with the $ p=0$ physical state.\par In general, these remnants of
higher string
excitations must be described by a topological field theory. It is
still to be
found.\par The above mechanism is known to operate in two dimensions
only. It
is unclear, whether it is generalizable to higher dimensions.\par To
summarize
this part - we probably need some self-tuning mechanism in our
universe and we
have few toy models for it, but a realistic theory is absent.
\section{The Dilaton}
The dilaton may play an important role in such a theory. It appears
naturally
in any string-theoretic description of gravity. Let us try to explain
its
origin in  the most general terms.\par  The Einstein theory describes
a
massless
tensor field $ G_{\mu\nu}\left( x \right)$ with the maximal possible
gauge
group, consisting of the diffeomorphisms on a given manifold, $
\left( DiffM
\right)$. This group removes all negative norm states of this field
and leaves
us with the physical gravitons.\par Let us now try a different logic.
Instead
of postulating the maximal gauge group, let us look for the minimal
one needed
for elimination of the negative norms.
Negative norms carried by tensor fields arise due to the Minkowskian
signature
of space - time. For example the vector potential, describing photons
has time
like components, corresponding  to oscillators with
negative norms. In QED gauge invariance guarantees that these ghost
states are
never emitted in real processes. The gauge  transformations in this
case
contain one arbitrary function. Hence, the ``size" of the gauge group
matches
the number  of  negative fields. In gravity the negative states are
produced by
the components $G_{0k}$ (where  $k$ is a space -like index). Hence we
need a
gauge group containing  $d-1$ arbitrary functions (where $d$ is a
dimension of
the manifold). It is easy to see that the simplest choice -the group
of volume
preserving  diffeomorphisms - eliminates all unwanted states. Since
it is
smaller than the Einstein group of all diffeomorphisms, it gives us
an extra
physical state -the dilaton. This is easy to see from the invariant
action:
\begin{equation}
S=\int{\left( a\left( \sqrt{G} \right)R+b\left( \sqrt{G} \right)
+\ldots\right)d^{d}x}
\end{equation}
Here $G=det(G_{\mu \nu)}$, while  $ a$ and $ b$ are some unknown
functions.
This action can be ``covariantized" by adding the dilaton field,
multiplying the
determinant of the metric. The above form, however is helpful for
finding  low
energy theorems since it relates dilaton emission to the reduced
covariance.
\par It is quite clear ,that in the theories with dilatons all
fundamental
couplings become dilaton - dependent. That means  that if the dilaton
remains
massless, it provides us with another self - tuning mechanism. A big
question
is whether this masslessness is consistent with the known cosmology
and the
equivalence principle. To answer this question one needs to know the
low
energy effective action to all orders in the string coupling. This is
another
big question. This action must be local, since all non-local terms
arising from
the massless exchanges contain high powers of  momenta. A possible
clue to the
problem is the restricted covariance mentioned above, since in string
theory it
appears through the anomaly in the complete covariance and thus might
be
tractable. I believe that the popular opinion that the dilaton must
be massive
is unwarranted before these big questions are answered.
\section{The Big Bang}
Talking about cosmology, I would like to mention the unusual picture
of Big
Bang in string theories, which has to modify our views on inflation.
Let us
consider Friedman anzatz:
\begin{equation}
ds^{2}\,=\, -dt^{2}+a^{2}\left( t \right)d\vec{n}^{2}
\end{equation}
Here $ \vec{n}^{2}=1$ represents a space element with constant
curvature.
\par In string theory one has to add the dilaton field  $ \Phi \left(
t
\right)$ and to consider (in the one loop approximation) the $
\beta=0$
equations [4]  which replace the Einstein equation. Just as in the
usual case,
it gives a Big Bang singularity,  represented by  the vanishing of $
a^{2}\left( t \right)$ at some moment  of time.\par In order to
understand the
origin of this result it is helpful to take a different view on the
string
theory in the background. Namely, let us introduce $ t=i\phi$ and
consider a
problem of the non-linear $ \sigma$- model, described by the  $
\vec{n}$- field
coupled to the two-dimensional gravitational field, described (in the
conformal
gauge) by the Liouville field $ \phi$.  In this interpretation, the
field $
\phi$ defines a scale, while $ a^{2}=a^{2}\left( \phi \right)$ is an
inverse
running coupling constant for this model. It is easy to write the
standard $
\beta$ - function equations for $ a^{2}\left( \phi \right)$ and the
dilaton $
\Phi\left( \phi \right)$ :
\begin{eqnarray}
a'a\Phi ' + a''a  + (N-2)(a')^{2} =N-2 \nonumber \\ (N-1){a'' \over
a} + \Phi
''=0
\end{eqnarray}

Now, the Big Bang singularity which appear in these equations is
nothing but
the standard logarithmic pole signifying asymptotic freedom. Of
course, this
pole is somewhat dressed by gravity. The theory of gravitational
dressing is a
fascinating subject by itself, which I hope to discuss elsewhere. For
the
present purposes it is sufficient to know that in one loop
approximation
gravity doesn't  introduce qualitative changes of the formulas  of
asymptotic
freedom. \par The undressed formula for $ a^{2}$ is given by [5] :
\begin{equation}
a^{2}\left( \lambda \right)= a^{2}\left( \lambda _{0} \right)-{N-2
\over
2\pi}\log{{\lambda \over \lambda _{0}}}
\end{equation}
\par Here $ \lambda$ is the scale and is  related to the Liouville
field. This
relation is somewhat complicated in general, but for large  $ \phi$
and $
\lambda$ it simplifies to:
\begin{equation}
\phi \propto \log{\lambda}
\end{equation}
 The solution of eq. (15) will give the same vanishing of $ a^{2}$ as
a
function of $\phi$ except that it will not be a simple zero in terms
of $
\phi$. \par Now, we come to the crucial point.\par As was explained
in [5] this
pole is just an artefact of the one-loop approximation. What happens
in reality
is that the mass gap is formed and the growth of interaction stops.
In terms of
$ a^{2}\left( \phi \right)$ it means that before the Big Bang we
enter a novel
regime with the massive $ \vec{n}$ - field. It seems that the finite
mass gap
of the $ \vec{n}$ - field implies that the metric tensor and
coordinates lose
their classical meaning. In a very early universe space-time does not
exist.
Instead we have some string Hilbert space description of the universe
which in
some approximation gives the metric tensor, but the approximation
breaks at the
logarithmic pole. No singularities are present and the ``big bang" was
not that
big after all. \par The physics is hidden in the amplitudes:
\begin{eqnarray}
A\left( l_{1}m_{1}\ldots l_{n}m_{n} \right)=<V_{l_{1}m_{1}}\ldots
V_{l_{n}m_{n}}>\nonumber \\V_{lm}=\int{d^{2}\xi Y_{lm}\left(
\vec{n}\left( \xi
\right) \right)}\psi _{l}\left( \phi \left( \xi \right) \right)
\end{eqnarray}
\par Here  $ Y_{lm}$ - are the spherical functions and  the Liouville
eigenfunctions $\psi_{l}$  are  determined from the conformal
invariance.\par
In the WKB limit it must be possible to reconstruct the space-time
metric from
the amplitudes (17), but beyond that we should view (17),  as a set
of numbers
characterizing the state of the universe, its whole history. \par
When passing
to Minkowskian regime, we have, perhaps, to change simultaneously a
sphere $
\vec{n}$ to hyperboloid in order to have vanishing of $ a^{2}\left( t
\right)$
at real time. But this requires further investigations.\par In any
case in this
unusual situation one has to reconsider all the standard problems
(horizon
etc.) which led to the inflation scenario. It is clear that the
argument that
the homogeneity of the Friedman universe requires acausal connections
at the
early times is not very restrictive in our picture. Indeed, these
early times
are described by the wave functional of the non-linear $ \sigma$ -
model, and
there are acausal connections of the Einstein-Podolsky -Rosen type.
May be
inflation (whatever it means in this context) is simply unnecessary,
although
the ``flatness" and the  ``monopoles" problems are still to be
resolved. \par Of
course, the above sketch is not a complete theory but a tempting
program  for
future research.
\section{Fuzzy strings and 3d Ising model}

 Some years ago I have suggested that the three-dimensional Ising
model can be
reduced to the theory of free strings [6].  This suggestion has been
further
developed in a number of papers [7]. My basic argument was the
following.
Consider first the  2d - Ising model. In this case we have the order
parameters  $
\sigma_{\vec{x}}$ and disorder parameters $ \left\{ \mu_{\vec{x}}
\right\}$  ,
each of which satisfy complicated equations. It is possible, however
to form
fermionic fields $ \psi_{a}\left( \vec{x} \right)$ where $ \vec{x}$
is a point
of the lattice and $ a=1,..4$ denotes one of the four possible
directions from
$ \vec{x}$ to the center of the adjacent plaquette.\par It is
common
knowledge by now that $ \psi_{a}\left( \vec{x} \right)$  satisfy
linear
equations, which in the continuous limit become 2d Dirac equations:
\begin{equation}
i \vec{\tau}{\partial \over \partial \vec{x}}{\chi}=m\chi
\end{equation}
  with $ \chi_{\alpha}\left( \vec{x} \right)$ , $ \alpha= 1,2$ being
$ spin {1
\over 2}$ component of $ \psi_{a}$ while $ spin {3 \over 2}$
component is not
propagating (see e.g. [8] for details).\par In three dimensions we
have order
variables $ \sigma_{\vec{x}\alpha}$ attached to each link (we treat
the model
as $ Z_{2}$ gauge theory) and the disorder variables placed at the
centers of
the cubes.  Let us consider the objects in the loop space :
\begin{equation}
\Psi_{a_{1}\ldots a_{L}}\left( C
\right)=\prod_{C}{\sigma_{\vec{x}\alpha}}\prod{\mu_{\hat{x_{a}}}}
\end{equation}
  (where $ C$ - is a closed loop of the length $L $ , and $ a=1,..4$
denotes
one of four possible cubes adjacent to the link $ \left( \vec{x}
\alpha\right)$
).\par It is easy to see that this object satisfies a  linear
equation
in the loop
space (modulo contact terms). The precise form of these equations is
given in [7].
They somewhat simplify in the hamiltonian version of this theory.
Namely, if,
by squeezing the lattice in the ``time" direction we pass to the
continuous time, we
obtain planar loops with arrows looking inside or outside the loop.
Correspondingly we introduce the object $ \Phi_{\alpha_{1}\ldots
\alpha_{L}}\left( C; t_{1}\ldots t_{L} \right)$ with $ C$  being  a
planar loop
of the length $ L$ while $ \alpha =1,2$ describes the direction of
the arrow.
By simple manipulations  one  gets the equations
\begin{equation}
{\partial \over \partial t_{s}}{\Phi \left( C \right)}=u\tau^{x}_{s}
\Phi\left(
C \right)-v\Phi\left( C+\tau^{z}_{s}\Pi_{s} \right)
\end{equation}
Here $ \tau -s$ are Pauli matrices acting on the index  number $ s$
,the symbol
in the last argument means that the link number $ s $ is replaced by
the
outside looking letter $ \Pi$ if $ \alpha_{s}=1$ and by the inside
looking $
\Pi$ otherwise.
In this form(4) is true only for convex contours. Otherwise, there
are some
local sign changes. Quantities $ u$ and $ v$ are related to the
parameters of
the Ising hamiltonian.\par Equations (20) are the counterpart of
the 2d
Ising equations. They have a very simple physical meaning. Namely they
imply that
each bit of the 3d Ising string moves as a spinning particle. To see
it ,
compare (20) with the lattice Dirac equation:
\begin{equation}
{\partial \over \partial t}{\psi}=u\tau^{x}\psi\left( x
\right)-v\psi\left(
x+\tau^{z} \right)
\end{equation}
Here again the last argument means the jump to the left or to the
right,
depending on the direction of spin.
\par In the 2d Ising case it is straightforward to take the continuum
limit of
(20) which gives the Dirac equation (18). \par However, any reliable
technique to
deal with the loop space equations (19) will perhaps appear in the
next
century. In the present paper I shall try to guess what will be the
result of
these future considerations. Of course, it is dangerously easy in
these
circumstances to chose an  inappropriate universality class, or in
plain
language,
to go astray. The only consolation in this case is that the
universality class
may be interesting enough by itself to deserve the study.\par It is
obvious,
first of all, that we are dealing with a string theory, which
carries spin
density on its world sheet and that the motion of the string is
correlated with
the directions of the spin i.e. we have spin-orbit interaction. Also,
as was
explained above, each bit of the string moves as in a plane
orthogonal to the
corresponding link as a two-dimensional fermion. Therefore in a
thoughtless
continuous limit (which would  work nicely for particles) we should
replace
(20) by the following:
\begin{equation}
\tau^{x}\left( s \right){\delta \Psi\left( C \right) \over  \delta
t\left( s
\right)}+\tau^{z}\left( s \right){\delta \Psi \left( C \right) \over
\delta
x_{\bot}\left( s \right)}=m\Psi \left( C \right)
\end{equation}
Here the sign $ \bot$ means the functional derivative in the
direction
orthogonal to  $ {d \over ds}{\vec{x}}$. \par If we wish this
equation to make
sense, we have to specify, what we mean by $ \vec{\tau}\left( s
\right)$, or
what is the string generalization of the Dirac matrices. One possible
answer to
this question is well known.It is beautiful construction of Ramond
[9]
and
Neveu-Schwartz [10]. These authors generalized the  Dirac equation by
introducing anti-commutativity condition for $s\neq s'$ among Dirac
matrices,
or , in other words, by replacing them by new fermion fields. That
lead to the
super-symmetric string action of NSR.  It seems, however, that
3D-Ising wave
equation (22) requires a  different generalization  of the Dirac
equation. \par
The reason is simple. If we take $ s $ and $ s'$ far apart, matrices
$
\tau\left( s \right)$ and  $ \tau\left( s' \right)$ become
independent and
commute, rather then anticommute. There are no Jordan - Wigner
factors in the
equation (22).\par After this is understood, it immediately comes to
mind to
replace  $ \vec{\tau}\left( s \right)$ by the SO(3) current algebra
with,
perhaps, a  central extension. \par As we shall see, this is close to
the truth
but not quite correct since there are complicated renormalizations.
The problem
now is to find a continuous string theory which incorporates the
field  $
\vec{\tau}$ and couples it to the string field $ \vec{x}$. If we
forget about  the $
\vec{x}$ - field for a moment, we can easily write the action, which
describes
fluctuations of $ \vec{\tau}$. It is well known [11], that the
one-dimensional
Heisenberg antiferromagnet is described by the $ \vec{n}$ - field
with the  $
\theta$ - term:
\begin{equation}
S_{1}=\int{({1 \over 2 \alpha}(\partial_{a}\vec{n})^{2}+ i \pi
\varepsilon^{ab}\vec{n}[
\partial_{a}\vec{n}\partial_{b}\vec{n}])d^{2}\xi}
\end{equation}
and the $ \vec{n }$ - field can be identified with the Heisenberg
spins
$\vec{\tau_{s}}$ The model is gapless and is exactly soluble by
different
methods. \par Now, at the critical point of the Ising model, equation
(22) can
be interpreted as an orthogonality of  $ \vec{\tau_{s}}$  to the
direction of
string propagation. A  natural covariant formulation of this
condition
is :
\begin{equation}
\vec{n}\left( \xi \right)\partial_{a}\vec{x}\left( \xi \right)=0
\end{equation}
\par or in other words, we require that the  spins,  living on the
string must
be orthogonal to the world sheet.\par This is not a correct effective
action
yet. As we will see in a moment, the constraint (24) does not survive
in
renormalization. Indeed, let us add to $ S_{1}$ a Lagrange multiplier
ensuring
the constraint (24):
\begin{equation}
S_{2}=\int{\lambda^{a}\left( \vec{n}\partial_{a}\vec{x}
\right)d^{2}\xi}
\end{equation}
One can compute then the induced action $S_{3}\left( \lambda^{a}
\right)$. It
can be done directly by expanding it in powers of $ \lambda$. Since $
\lambda$
has dimension one, we expect and get logarithmic counterterms to the
induced
action. They have the form

\begin{equation}
S_ {3} \propto log\Lambda\int{\left( \lambda^{a} \right)^{2}d^{2}\xi}
 \end{equation}
As usual, it means that to keep the theory renormalizible one has to
introduce
a new coupling constant in front of (26) and to examine the  two
couplings
renormalization group. If we do this and integrate out the field $
\lambda$, we
obtain:
 \begin{equation}
S_{4}=\,f\, \int{\left( \vec{n}\partial_{a}
\vec{x}\right)^{2}d^{2}\xi}
 \end{equation}
where $ f $ is a new coupling. The geometrical meaning of the
replacement of
the constraint (24) by the action (27) is transparent : as we
renormalize and
go to the block-variables the constraint gets smeared and what was
the normal
becomes a tilted vector on the surface. We can now write the
effective action,
describing the string with spin density and spin -orbit interaction:
\begin{eqnarray}
S=\int{\left( {1 \over
2}\sqrt{g}g^{ab}\partial_{a}\vec{x}\partial_{b}\vec{x}
+\mu\sqrt{g}+{1 \over 2\alpha} \sqrt{g}g^{ab}\partial_{a}
\vec{n}\partial_{b}
\vec{n}+f\sqrt{g}g^{ab}\left( \vec{n}\partial_{a}\vec{x}
\right)\left(
\vec{n}\partial_{b} \vec{x} \right) \right) d^{2}\xi} \nonumber \\
+i \pi \int
\varepsilon^{ab}\vec{n}[\partial_{a}\vec{n}\partial_{b}\vec{n}]
d^{2}\xi
 \,\,\,\,\,\,\,\,\,\,\,\,\,\,\,\,\,\,\,\,\,\,\,\,\,\,\,\,\,\,\,\,\,\,
\end{eqnarray}
The last term in this expression is not an Euler character of the
surface,
since $ \vec{n}$ is not an exact normal. One can say that
renormalization
destroys Gauss-Bonnet theorem.\par An important feature of the action
(28) is
its renormalizability. There are highly non-trivial trajectories of
all
couplings involved, which up to now I explored only in the  one loop
approximation. However, the theory may be exactly solvable. Indeed,
its $ ``n"$
- part can be reduced to the $ k=1$  Wess-Zumino-Novikov-Witten
model,
and as
expected represents Kac-Moody currents, coupled to the world sheet.
Of course,
one should also couple it to the 2d gravity described by $ g_{ab}$,
so all in
all it is not an easy problem. But if solved, it has chances to
describe 3d
Ising model and 3d critical phenomena in general.
\section{The QCD string}
 From the  very  early years of QCD, on the basis of $ {1 \over N}$
expansion
[12], and strong coupling expansion [13] it was suspected that it is
somehow
related to string theory. Later  equations in the loop space,
satisfied by the
phase factors have been examined [6, 14, 15]. I had a hope to find
some string
lagrangian, such that the sum over surfaces satisfied these loop
equations,
much in the same way in which sums over paths satisfied the
Klein-Gordon
equations. \par In other words, the hope was for an  exact duality
between  the
gauge field representation and the  string representation of QCD. In
spite
of many
subsequent efforts [ 16, 17] the problem remains unsolved.\par In
this section
I will try to summarize were we stand now. The main object in the
theory is the
phase factor (the Wilson loop):
 \begin{equation}
\Phi\left( C \right)=<TrP\exp{\oint_{C}{A_{\mu}dx^{\mu}}}>
\end{equation}
where $  P $ - is an ordering sign and the brackets mean averaging in
QCD
vacuum. This quantity, together with more complicated correlations:
\begin{equation}
\Phi\left( C_{1}\ldots C_{n} \right)=<
TrP\exp{\oint_{C_{1}}{A_{\mu}dx^{\mu}}\ldots
TrP\exp{\oint_{C_{n}}}{A_{\mu}dx^{\mu}}}>
\end{equation}
satisfy the chain equations in  loop space. In the large $ N$
limit  we have
a relation:
\begin{equation}
\Phi\left( C_{1} \ldots C_{n}\right) \approx\Phi\left( C_{1}
\right)\ldots
\Phi\left( C_{n} \right)
\end{equation}
and hence one obtains a closed equation for $ \Phi\left( C \right)$.
Let us
explain the structure of this equation and its possible
solutions.\par The
basic idea is to find such an operator in the loop space, that being
applied to
$ \Phi\left( C \right)$ it will give on the right hand side an
expression
proportional to the Yang-Mills equations of motion. It is easy to
see, that the
required operator has the form:
\begin{equation}
{\partial^{2} \over \partial x^{2}\left( s
\right)}=\lim_{a\rightarrow
0}{\int_{-a}^{a}{dt{\delta ^{2}\over \delta x\left( s+{t \over 2}
\right)\delta
x\left( s-{t \over 2} \right)}}}
\end{equation}
and that the equation for $ \Phi\left( C \right)$ has the form:
\begin{equation}
{\partial^{2} \over \partial x^{2}\left( s \right)}\Phi \left( C
\right)=0\,
(\mbox{modulo contact terms})
\end{equation}
where the contact terms, described in the old papers [16, 17  ] are
nonzero
only for  self-intersecting loops. \par  Our aim is to find sum
over
surfaces representation which satisfies  equations (33), in the same
way , in
which Feynman's sums over paths satisfy the Klein - Gordon
equation.\par  At
this point we see the problem. All conventional string theories have
string
wave equations,  which contain an operator $ {\delta^{2} \over \delta
x^{2}\left( s \right)}$  instead of (32). The reason for that is the
presence
of the  $ (\partial x )^{2}$ term in the string action, which is
essentially
the Nambu term. This term, containing the square of the time
derivative of $ x$
unavoidably leads to the second order wave equation. At the same time
the
operator (32) is of the first order. Indeed, for any two functionals,
$ A$ and
$ B$ we have:
\begin{equation}
{\partial^{2} \over \partial x^{2}\left( s \right)}\left( AB
\right)=A{\partial^{2} \over \partial x^{2}\left( s
\right)}B+({\partial^{2}
\over \partial x^{2}\left( s \right)}A)B
\end{equation}
\par We come to the conclusion that we have to  find a string theory
in which
the Nambu term is absent or irrelevant, and try to adjust the action
so that
the wave operator takes the form (33). The second requirement of
course is the
correct form of the contact term. In what follows, I will attempt to
resolve
the first part, while the second one remains unclear.\par In the
standard
string theory we often introduce background fields -the metric $
G_{\mu\nu}$,
the antisymmetric field $ B_{\mu\nu}$, etc. It is clear from the
above that in
order to avoid the problem with the second derivative we have to make
the
following radical assumption about the background:
\begin{equation}
G_{\mu\nu}=0,
\end{equation}
while retaining the $ B_{\mu\nu}$ term, which contains the time
derivatives in
the first power.\par
Consider the following functional integral:
\begin{equation}
Z=\int{d\mu\left( B \right)\int{Dx\left( \xi
\right)\exp{\int{d^{2}\xi
B_{\mu\nu}\left( x\left( \xi \right)
\right)\varepsilon^{ab}\partial_{a}x^{\mu}\partial_{b}x^{\nu}}}}}
\end{equation}
Here $ Z$ is a partition function, $ x\left( \xi \right)$ describes a
closed
surface, and $ d\mu \left( B \right)$ is an unknown measure for the
antisymmetric field. In order to use the loop equation (33) one has
to consider
 an analogous expression for the surfaces with the boundary. Such
expressions
are even more difficult to define in the precise sense.
However if  one has the audacity to disregard multiple problems  with the
regularization, the result would be:
\begin{equation}
{\partial^{2} \over \partial x^{2}\left( s \right)}\Phi \left( C
\right)=<\partial_{\mu}B_{\mu\nu}\left( x\left( s \right)
\right)\dot{x}^{\nu}\left( s \right)>
\end{equation}
It is easy to adjust the B-measure so that the  RHS of (37) is zero.
This
happens to be the case for the Gaussian measures. Unfortunately,
under these
naive  operations the contact terms  don't  come out right and the
problem
remains unsolved. The only additional comment which I can make
concerns the
relation between the above models and the theory of  ``rigid strings"
[ 18]. For
the gaussian B-fields they are connected via the formula:
\begin{equation}
\int{d\sigma^{\mu\nu}\left( \xi
\right)d\sigma^{\mu\nu}\left(\eta\right)\delta
\left( x\left( \xi \right)-x\left( \eta \right)
\right)}=\,\mbox{const}A+\int{K^{2}d^{2}\xi}
\end{equation}
 Here $ A$ is the area of the surface, $ K$ is an extrinsic curvature
and :
\begin{equation}
d\sigma^{\mu\nu}=\varepsilon^{ab}\partial_{a}x^{\mu}\partial_{b}x^{\nu}d^{2}\xi
\end{equation}
The constant in (38) is  quadratically divergent and perhaps should
be set to
zero. Most probably, some type of the  $ \vartheta$ - term will be
needed for
the theory to make sense. These  terms for rigid strings serve to
restore
unitarity, apparently broken by the higher derivatives [ 18] . One of
the ways
to check the above representation may be to compare it with the
recent results
in two-dimensional QCD [19] which suggest ``stringy" structure in this
case.
Encouraging signs for such a comparison are irrelevance of the folds
and
volume- preserving diffeomorphism invariance in both cases. But much
more still
has to be done.
 \section{Scale dependence in quantum gravity}
One of the most important tools of quantum field theory is  the
renormalization
group. It tells us how different quantities  depend on the overall
scale. In
order to approach the problems described in the preceding sections we
need to
develop  a similar apparatus in the presence of  the two-dimensional
quantum
gravity. Here again we know only few elementary facts, while the
problem as a
whole is still  mysterious. \par First of all, there is some
conceptual puzzle
in the question of scale dependence. It  is well known  that in the
usual
theories scale transformations are generated by the trace of the
energy-
momentum tensor, $ T_{aa}$. At the same time, in any theory of
gravity, all
components of the energy - momentum tensor are  zero or , if one
wishes to fix
a gauge, are BRST commutators.\par Let us first resolve this puzzle
in the most
general terms. The apparent vanishing of the energy- momentum tensor
follows
from the relation:
\begin{equation}
\int{Dg_{ab}{\delta \over \delta g_{ab}\left( \xi
\right)}\exp{(-S\left( g_{ab}
\right))}}=0 (?)
\end{equation}
Here  $ S$ is the action, the $ g_{ab}$ is a two-dimensional metric.
We have
used  a somewhat symbolic notations in which we don't fix the gauge
explicitly.
More careful definitions wouldn't change the result and hence are not
needed.
Since:
\begin{equation}
T^{ab}={\delta S \over \delta  g_{ab}}
\end{equation}
we come to the above mentioned conclusion, that all matrix elements
of the
energy- momentum tensor are zero. This conclusion is wrong. The
source of the
error is the boundary in the space of all geometries, which produces
non-zero
boundary contribution to the integral (40).
Indeed, if the metric $ g_{ab}$ has two eigenvalues, $ \lambda_{1,2}$
we must
constrain the integration region in (40) by the condition:
\begin{equation}
\lambda_{1,2}\geq 0
\end{equation}
An interesting conclusion which we can draw from this consideration
is that the
scale dependence  or the  $ \beta$ function in the theory of gravity
must be
determined by degenerate geometries, say by  pinched spheres if
the overall
topology is spherical. \par Let us first demonstrate how this idea
works in the
most trivial case - the theory of random paths. Consider the action
for the
path in the background field $ G_{\mu\nu}$:
\begin{equation}
S=\int{G_{\mu\nu}\left( x\left( \tau \right)
\right)\dot{x}^{\mu}\dot{x}^{\nu}h^{-1}\left( \tau \right)d\tau}
\end{equation}
Here $ h\left( \tau \right)$ is a one dimensional metric on the path.
Let us
consider now  the partition function, defined by:
\begin{equation}
Z=\int{(\exp{-S}) Dh\left( \tau \right)Dx\left( \tau \right)}
\end{equation}
The scales in the ambient space and in the internal space are
related. Take the
Weyl variation of $ Z$:
\begin{equation}
G_{\mu\nu}{\delta Z \over \delta G_{\mu\nu}}=<\int{T\left( \tau
\right)\delta\left( x-x\left( \tau \right) \right)h\left( \tau
\right)d\tau}>
\end{equation}
Here $ T\left( \tau \right) =
h^{-2}G_{\mu\nu}\dot{x}^{\mu}\dot{x}^{\nu}$ is
the one dimensional metric tensor. Since we have to integrate over
the metrics
$ h\left( \tau \right)$ the naive expectation would be that  the RHS
of (45) is
zero. In fact it  is described by the boundary in the space of  all
1d
geometries, which consists of the loops  of zero (or better to say
infinitesimal) length. These loops contribute to the RHS of (45) a
term,
which is  local in $ G_{\mu\nu}$ and is the standard conformal
anomaly, derived
in a nonstandard way. The locality of this term  (while $ Z$ itself
is highly
non-local) is the reflection of its  ``zero loop" origin.\par When we
attempt to
generalize this consideration  to strings many complications arise.
It is not
easy to describe  the pinched geometries, mentioned above.
Nevertheless some
general statements can be made.\par Let us consider a partition
function  for
non- critical string in some background graviton -dilaton field, $
Z=Z\left(
G_{\mu\nu}, \Phi \right)$. Let us stress that in this case, unlike
critical
strings, the background is not fixed by any equations of motion and
can be
chosen more or less arbitrarily. The functional $ Z$ is given by a
formula,
similar to (44). In order to find its scale dependence, we have to
devise such
a transformation of the space-time quantities that it  is reduced to
the
energy- momentum tensor on the world sheet, as was the case with
(45). This is
easily achieved if instead of the Weyl rescaling  we consider the
following
change in the target space, which we will call the $ \beta$ -
symmetry:
\begin{equation}
\delta G_{\mu\nu}\left( x \right)=\varepsilon\left( x
\right)\beta_{\mu\nu}\left( x \right) ; \delta \Phi\left( x \right)=
\varepsilon\left( x \right)\beta^{\Phi}\left( x \right)
\end{equation}
It is clear now that under this symmetry the partition function will
change in
the following way:
\begin{eqnarray}
\beta_{\mu\nu}\left( x \right){\delta Z \over \delta
G_{\mu\nu}}+\beta^{\Phi}\left( x \right){\delta Z \over \delta
\Phi}=<\int{T_{a}^{a}}\left(\xi \right)\delta \left( x-x\left( \xi
\right)
\right)\sqrt{g}d^{2}\xi> =\mbox{anomaly} \nonumber \\
\mbox{anomaly}\,=\,\Gamma^{IJ}\left( \lambda^{K}\left( x \right)
\right){\delta Z
\over \delta \lambda^{I}}{\delta Z \over \delta \lambda^{J}}+f\left(
G^{\mu\nu}\left( x \right) ,\Phi\left( x \right) \ldots\right)
\end{eqnarray}
Here the fields  $ \lambda^{I}$  correspond to the different string
states,
the $ \Gamma^{IJ}$ is some metric in the space of these states and $
f$ is some
local function of fields corresponding to these states; the world
sheet is
assumed to have spherical topology.\par This formula follows from the
fact that
the RHS of (47) is entirely dominated by the degenerate metrics. The
first term
corresponds to the pinching of the world sheet at  a given point,
while the
second local term comes from the complete collapse of the world
sheet.
Unfortunately  it is not known in general how to calculate the
quantities
entering in (47).  Nevertheless, this equation has an interesting
interpretation from the point of view of critical string in one extra
dimension
(which, as usual, is interpreted as the Liouville coordinate. In this
case we
introduce background fields $ G_{\mu\nu}\left( x, \phi \right)$, $
\Phi\left(
x, \phi \right)$ etc. This time those fields can not be arbitrary.
They must
satisfy equations of motion which follow from the vanishing of all $
\beta$ -
functions. It is well known that these equations follow from the
least action
principle  for a certain effective action $ W$ [4]. It can be shown
that the
background fields, introduced in the non-critical context are simply
initial
data for  the critical background:
\begin{equation}
G_{\mu\nu}\left( x \right)=G_{\mu\nu}\left( x,\phi_{0} \right) ;
\ldots
\end{equation}
for some fixed $ \phi_{0}$. Moreover the partition function $ Z$ is
simply
the value  of the classical action $ W$ on the solution with the
initial values
(48):
\begin{equation}
Z\,=\,\mbox{min}W
\end{equation}
With this understanding, one can interpret (47) as a Hamilton -Jacobi
equation
for the classical action. Perhaps this is a simplest way to find the
contribution of the pinched spheres, at least in a one loop
approximation. Still
it is very desirable to have direct methods for the computations.
\par It is
clear that we need a  much better understanding  of the
renormalization
group in
the presence of gravity. I believe that gravity should actually
simplify the
structure of the renormalization, since it  decreases the number of
physical
states by means of decoupling of the descendant operators.  It can be
hoped
that there are some very general rules governing  the scale
dependence. These
rules are still to be found. Here I shall quote  some partial results
obtained
together with I. Klebanov and I. Kogan. We showed that  in the  one
loop
approximation  all the  $ \beta$ functions get changed by gravity,
acquiring
an extra factor  $ {k+2 \over k+1}$ where $ k$ is the central charge
of
the
gravitational $ SL\left( 2 \right)$ algebra. What happens in the
higher orders
is a fascinating question. \section{Conclusions}
It is obvious that we have a lot of interesting, difficult and
important work
ahead of us.
 \section{Acknowledgements}
I am very grateful to the organizers of Les Houches school for  their
successful efforts to create a very stimulating  atmosphere, which
mixed so
well with the mountain air.
This work was partially supported by the National Science Foundation
under
contract PHYS-90-21984.

\eop
\appendix{REFERENCES\par [1] L. Okun Usp. Fiz. Nauk 161, 9, (1991),
177\par [2]
S. Coleman Nucl. Phys. B310(1988)64\par[3] A. Polyakov Sov. Phys.
Usp. 25
(1982) 187
\par [4]C.
Callan et al. Nucl. Phys. B262, (1985) 593\par[5] A. Polyakov Phys.
Lett.59B
(1975) 80\par [6] A. Polyakov Phys. Lett. 82B (1979) 247\par [7] E.
Fradkin
M.Srednicky L. Susskind Phys. Rev.D21 (1980) 2885;
\par ~~~~C. Itzykson, Nucl. Phys. { B210}, 477 (1982);
\par ~~~~A. Sedrakyan
Phys. Lett.
137B (1984) 397;
\par    ~~~~A. Casher, D. F\oe rster and P. Windey, Nucl. Phys. { B251}, 29
(1985);
\par    ~~~~Vl. Dotsenko and A. Polyakov, {\it in} Advanced Studies in Pure
    Math. { 15} (1987).
\par [8] A. Polyakov ``Gauge fields and Strings"
Harwood Academic
Publishers (1987)\par[9] P. Ramond Phys. Rev.D3(1971) 2415\par [10]
A. Neveu J.
Schwarz Nucl. Phys. B31 (1971) 86\par [11] F.D.M. Haldane Phys. Lett.
 93 A (1983)  464.
\par[12] G. t' Hooft Nucl. Phys. B72 (1974) 461 \par[13] K. Wilson
Phys. Rev.
D8(1974)2445\par [14] J. Gervais A. Neveu Phys. Lett. 80B (1979)
255\par [15]
Y. Nambu Phys. Lett. 80B (1979) 372\par [16] Yu. Makeenko, A Migdal
Nucl. Phys.
B188(1981) 269\par [17] A. Polyakov Nucl. Phys. B164 (1980) 171\par
[18] A.
Polyakov Nucl. Phys. B268 (1986) 406 \par [19] D. Gross  W. Taylor
Princeton
preprint PUPT-1382 (1992)}

\end{document}